\documentstyle[preprint,prc,aps]{revtex}
\begin{document}
\draft
\title{HEAVY-FLAVORED STRANGE PENTAQUARK SEEMS NOT TO EXIST}
\author{M.Shmatikov}
\address{Permanent address: Russian Research Center "Kurchatov\\
Institute", 123182 Moscow, Russia}
\maketitle

\begin{abstract}
Possible existence of a molecular-type heavy-flavored strange pentaquark
is considered. No narrow state of such a type is shown to exist.
\end{abstract}


\tighten
\section{Introduction}
All the observable hadrons can be successfully classified at present as $%
\bar qq$ or $qqq$ configurations. The challenging question of the existence
of exotics remains unanswered. One of such an exotic states is the
pentaquark: baryon containing 5 quarks $P_{\bar Q}\equiv \bar Qqqqq$ where $%
Q\,(=b,c)$ is a heavy quark and $q\,(=u,d,s)$ is a light one. Theoretical
search for a pentaquark yielded controversial results. Its existence has
been claimed in various approaches. Investigation of the pentaquark
existence in the constituent quark model resulted in a confirmative
conclusion provided 1) one of the light quarks is a strange one and 2) the
binding color forces are assumed to be flavor-independent \cite{richard},
\cite{lipkin}. Stated differently in the limit of ``flavor-blind'' color
forces the $P_{\bar Qs}$ is expected to exist. However, the realistic
breaking of the $SU(3)_F$ symmetry of QCD forces and, even more important,
the motion of quarks being taken into account, the pentaquark proved to be
unstable with respect to strong decay \cite{fleck},\cite{zouzou}.

Alternative approach to the issue is to consider the pentaquark as a bound
state of Skyrme-type soliton (representing the nucleon) and a heavy meson.
This model has been applied in the nonstrange sector only and the results
obtained are far from being decisive. The bound state was shown to exist in
both $b$- and $c$-sectors (corresponding to the $P_{\bar b}$ and $P_{\bar c}$
type pentaquarks) in the limit of the infinite heavy-quark mass \cite{riska}%
. At the same time corrections due to the finite mass of the heavy quark may
make the pentaquarks states unbound \cite{Oh}.

Finally, a molecular-type pentaquark $P_{\bar b}$ state is predicted \cite
{sh}. It is bound by the long-range one-pion-exchange forces ensuring its
genuine molecular structure. However, these forces are not strong enough to
bind a $P_{\bar c}$ pentaquark. In view of this controversial situation it
is of interest to explore whether the molecular-type $P_{\bar Qs}$
pentaquark exists.

\section{Quantum numbers of a molecular-type strange heavy-flavored
pentaquark}

Molecular-type hadron can be characterized by its particle content. The
nonstrange $P_{\bar b}$ pentaquark is a loosely bound state of the $\bar B%
^{*}N$ system.. ``Naive'' extension of this approach to the system
containing the nucleon and the bottom-flavored strange meson $\bar B_s$($%
\equiv (\bar bs)$) fails: because of the $\bar B_s$-meson zero isospin it
does not couple with the $\pi $-meson. The viable alternative is a molecular
state comprising the (nonstrange) heavy quark $H^{*}$\footnote{%
Since it does not result in confusion we hereafter will denote for
notational brevity the heavy vector meson as $H$.} and a hyperon $Y$. Two
possibilities are conceivable depending on the total isospin of the system $T
$. Consider first the $T=3/2$ case. Then the pentaquark-to-be necessarily
has the $H-\Sigma $ particle content. The ``effective'' strength constant of
the driving one-pion-exchange potential in the $T=3/2$ channel equals then $%
\nu $$=f_{\Sigma \Sigma \pi }\cdot f_{HH\pi }\cdot \gamma $, where $\gamma $
is the spin factor. In the limit of the $SU(3)$ symmetry $f_{\Sigma \Sigma
\pi }=2\alpha _P\,f_{NN\pi }$, where $f_{NN\pi }$ is the constant of the $%
NN\pi $ coupling and $\alpha _P$ is the $F/(F+D)$ ratio for the pseudoscalar
mesons. The $\pi $-meson is known to couple to the isovector particles only
and, hence, it is uncoupled from the isoscalar heavy quarks. Introducing the
constant $g$ of the $\pi $-meson coupling to the constituent light quark we
arrive at $f_{HH\pi }=g$. The coupling constant $g$ is related also to the $%
NN\pi $ coupling constant: $g=3/5\,f_{NN\pi }$. Combining all the enlisted
factors we get the following expression for the ``effective'' strength
constant of the one-pion-exchange in the $T=3/2$ channel :
\[
\nu _{H\Sigma }=6/5\,\alpha _P\gamma \,\,\,f_{NN\pi }^2
\]
It is instructive to compare this strength constant to its counterpart $\nu $%
$_{NN}$in the $NN$ system . The latter with the account of the spin-isospin
factors reads
\[
\nu _{NN}=3\,f_{NN\pi }^2
\]
The ratio of $\nu $$_{H\Sigma }$ and $\nu $$_{NN}$ strength constants is
\[
\nu _{H\Sigma }/\nu _{NN}=2/5\alpha _P\gamma
\]
The $\alpha _P$ factor in the limit of the $SU(6)$ symmetry equals 0.4. Its
value obtained in the Nijmegen potential equals $\alpha _P=0.485$ \cite{Nij}%
. The value of the spin factor $\gamma $ depends on the spin and orbital
momenta of the considered state. For the most favorable $^2S_{1/2}-^4D_{1/2}$
configuration (see \cite{sh}) it is $\gamma =2$. Thus we arrive at
\[
\nu _{H\Sigma }/\nu _{NN}=4/5\alpha _P
\]
showing that $\nu $$_{H\Sigma }\approx 1/3\,\nu _{NN}$. One more factor
which is important for binding is the reduced mass of the system. The ratio
of reduced masses for the $H-\Sigma $ and $NN$ systems equals
\[
\frac{m_{H\Sigma }}{m_{NN}}=2\,\frac{m_\Sigma }{m_N}\,\frac 1{1+m_\Sigma /m_H%
}
\]
This ratio equals 1.55 (2.08) in the case of $D$- ($B$-) heavy meson. The
increased reduced mass of the $H-\Sigma $ system as compared to the $NN$
case still cannot overbalance the smallness of the effective strength
constant $\nu $$_{H\Sigma }$ allowing to conclude safely that the
molecular-type $H-\Sigma $ pentaquark with the $T=3/2$ total isospin does
not exist.

Alternative possibility is the $H-Y$ molecular state with the total isospin $%
T=1/2$. This case comprises two coupled channels: $H-\Lambda $ and $H-\Sigma
$. As it was stated above the configuration which is the most favorable from
the point of view of binding by the one-pion-exchange has the $J^\pi
=1/2^{-} $ quantum numbers corresponding to the $^2S_{1/2}-^4D_{1/2}$
coupled channels. Thus we arrive at the system of 4 coupled Schr\"odinger
equations which can be written in the matrix form
\begin{equation}  \label{sys}
\frac{d^2\phi }{dr^2}={\cal K}^2\,\phi +2mV\phi
\end{equation}
where $\phi $ is the column of wave functions with the following numeration
of channels: (1) $H-\Lambda $ ($S$-wave); (2) $H-\Lambda $ ($D$-wave); (3) $%
H-\Sigma $ ($S$-wave); (4) $H-\Sigma $ ($D$-wave) and ${\cal K}^2\ $ is the
diagonal $4\times 4$ matrix of c.m.s. momenta squared (plus the matrix of
centrifugal potentials):
\begin{equation}  \label{mom}
{\cal K}^2=\left[
\begin{array}{cccc}
\kappa _{\Lambda H}^2 &  &  &  \\
& \kappa _{\Lambda H}^2+6/r^2 &  &  \\
&  & \kappa _{\Sigma H}^2 &  \\
&  &  & \kappa _{\Sigma H}^2+6/r^2
\end{array}
\right]
\end{equation}
The c.m.s. momenta squared in eq.(\ref{mom}) for the kinetic energy $E$ in
the $H\Lambda $ channel are equal to $\kappa _{\Lambda H}^2=2m_{\Lambda
H}\cdot E$ and $\kappa _{\Sigma H}^2=2m_{\Sigma H}(E+\Delta m)$, where $%
\Delta m$ is the mass difference of the $\Sigma $- and $\Lambda $-hyperons ($%
\Delta m=m_\Sigma -m_\Lambda \approx 76$ MeV) and $m_{YH}$ ($Y=\Lambda
,\Sigma $) is the reduced mass in the corresponding channel. In eq.(\ref{mom}%
) $m$ is the diagonal $4\times 4$ matrix with the $m_{\Lambda H}$ and $%
m_{\Sigma H\text{ }}$elements. Finally, $V$ is the matrix of driving forces
reading:
\begin{equation}  \label{sch}
V^\pi =V_0\left[ T\otimes C_s\,\tilde y_0(m_\pi r)+T\otimes C_t\,\tilde y%
_2(m_\pi r)\right]
\end{equation}
where $V_0$ is the potential unit introduced in \cite{To}
\begin{equation}  \label{ps}
V_0=\frac{m_\pi ^3}{12\pi }\,\frac{g^2}{f_\pi ^2}
\end{equation}
where $g$ is the constant of the $\pi -$meson coupling to the constituent
quark and $f_\pi $ is the constant of the $\pi -$meson decay ( $f_\pi
\approx 132$ MeV). In eq.(\ref{sch}) $\tilde y_l$ ($l=0,2$) are the
regularized at small $x$'s radial functions
\begin{equation}  \label{yra}
\,\tilde y_0(x)=y_0(x)-y_0(\lambda x)-(\lambda -1)\exp (-\lambda x)\medskip\
,
\end{equation}
and
\begin{equation}  \label{yrb}
\,\tilde y_2(x)=y_2(x)-\lambda ^3y_2(\Lambda x)-\exp (-\lambda x)\,\frac{%
(\lambda ^2-1)(\lambda x+1)}{2x}
\end{equation}
where $y_{0,2}(x)$ are the spherical modified Bessel functions
\[
\begin{array}{cc}
\,y_0(x)=\exp (-x)/x; & \,y_2=\exp (-x)/x\cdot (1+3/x+3/x^2)
\end{array}
\]
and $\lambda =$$\Lambda /m_\pi $, with $\Lambda $ being a cut-off parameter
(see below). The $C_{s(t)}$ matrix contains spin coefficients for the
central (tensor) components of the one-pion-exchange forces:
\[
\begin{array}{cc}
C_s=\left[
\begin{array}{cc}
-2 & 0 \\
0 & 1
\end{array}
\right] ; & C_t=\left[
\begin{array}{cc}
0 & -1 \\
-1 & \sqrt{2}
\end{array}
\right]
\end{array}
\]
Finally, $T$ is the matrix of isospin coefficients which in the limit of the
$SU(3)$ symmetry limit reads
\begin{equation}  \label{tm}
T=\;
\begin{tabular}{c|c|c}
& $H-\Lambda$ & $H-\Sigma$ \\ \hline
$H-\Lambda$ & 0 & $-10/3\:(1-\alpha_P)$ \\
$H-\Sigma$ & $-10/3\:(1-\alpha_P)$ & $-20/3\:\alpha_P$ \\
&  &
\end{tabular}
\end{equation}
where $\alpha _P$ is defined above.

\section{Binding of a molecular-type pentaquark}

Complexity of the system under consideration make application of numerical
methods mandatory. Before proceeding to the numerical solution of the system
(\ref{sys}) let us consider either conditions for emerging of a loosely
bound state are favorable. To this end we apply the criterion for the
appearance of a zero-energy level in a system where a central force is
operative \cite{Blatt}. For the Yukawa forces under consideration the
parameter to be investigated reads
\[
s=0.5953\cdot 2\bar m\gamma \,\frac{V_0}{m_\pi ^2}
\]
where $\bar m$ is the reduced mass of the system and $\gamma $ is the
spin-isospin factor. The only component of the system under consideration
where the Yukawa forces are operative is the $S$-wave $H-\Sigma $ channel.
To make conclusions more transparent we compare the $s_{\Sigma H}$ value to
the (reference) value of its $NN$ counterpart $s_{NN}$:
\begin{equation}
\frac{s_{\Sigma H}}{s_{NN}}=\frac{\gamma _{\Sigma H}}{\gamma _{NN}}\cdot 2%
\frac{m_\Sigma }{m_N}\cdot \frac 1{1+m_\Sigma /m_H}  \label{rat}
\end{equation}
The spin-isospin $\gamma _{\Sigma H}$ factor equals the product of
corresponding elements of the $C_s$ and $T$ matrices:
\[
\gamma _{\Sigma H}=2\cdot 20/3\cdot \alpha _P
\]
In the same notations the $\gamma _{NN}$ equals
\[
\gamma _{NN}=25/3
\]
Substituting these expressions into eq.(\ref{rat}) we arrive at
\[
\frac{s_{\Sigma H}}{s_{NN}}\approx \frac{4\alpha _P}{1+m_\Sigma /m_H}
\]
This ratio depends upon the mass of the heavy meson $m_H$ and for the case
of the $D$- ($B$-) meson it is
\begin{equation}
s_{\Sigma H}\approx s_{NN}\cdot 1.0\;(1.3)  \label{cri}
\end{equation}
The value of $s_{NN}\approx 0.33$ is known to be smaller than that required
for the emergence of a zero-energy level ($s\geq 1$) \cite{Blatt},
nevertheless existence of the deuteron and the virtual bound state in the $%
^1S_0$ channel suggests it as a benchmark for the appearance of a loosely
bound state. Inspection of (\ref{cri}) shows that the conditions for the
binding of the $H-\Sigma $ pair are at least not worse than in the $NN$
system.. It should be emphasized, however, that this conclusion should be
considered rather as guide. Indeed, the pentaquark-to-be spends part of the
time in the $H-\Lambda $ channel where the interaction is absent. The
resulting attraction is weakened and the existence of a bound state can be
explored by the numerical solution of the system (\ref{sys}).

The equations to be solved involve two parameters controlling the strength
of the interaction. First, it is the constant $g$ of the $\pi $-meson
coupling to the constituent quark which determines the value of the
potential unit (\ref{ps}). The value of $g$ can be extracted from the
well-known constant of the pseudovector $NN\pi $ coupling:
\begin{equation}  \label{gc}
g=3/5\,f_{NN\pi }
\end{equation}
where $\,f_{NN\pi }^2/4\pi =0.08$. However, the coupling of the $\pi $-meson
to a constituent quark bound in the heavy $H$-meson may be altered by the
presence of the heavy quark. Another parameter affecting the strength of the
one-pion exchange is the formfactor cut-off $\Lambda $ (see (\ref{yra}) and (%
\ref{yrb})). The larger is $\Lambda $ the more are potentials peaked at
small $x$'s and the stronger is the resulting potential. The value of $%
\Lambda $ is poorly known: in different models of the $NN$ interaction it
varies from 0.8 GeV to 1.5 GeV. Note, however, that variations of $g$ and $%
\Lambda $ affect the strength of the interaction in the same direction: when
either of these parameter increases the one-pion-exchange potential becomes
more strong.

The system (\ref{sys}) up to apparent modifications describes as well the $NN
$ interaction in the $^3S_1-^3D_1$ state. In this case the isotopic $T$
matrix (\ref{tm}) becomes a diagonal one with all the matrix elements equal
to -25/3, and the system (\ref{sys}) splits into two identical $2\times 2$
subsystems. The value of $g$ which follows from (\ref{gc}) equals $g\approx
0.6$ and the corresponding value of the potential unit $V_0\approx 1.3$ MeV
\cite{To}. The deuteron binding energy $\epsilon _D\simeq 2.2$ MeV is
reproduced with $g$ value increased by 15\% (30\%) for the values of the
cut-off parameter $\Lambda =1500$ MeV (1200 MeV). Note that these values of
the potential-strength parameters are somewhat higher than those obtained in
\cite{To}.

With the same set of parameters we solved eqs. (\ref{sys}) for the
considered $B\Lambda -B\Sigma $ system to find out the existence of a
loosely bound state. According to expectations the binding energy increases
with both $g$ and $\Lambda $ growing (see fig.1). Due to the small value of
binding energy the mass of the state is close to the sum of masses of its
components: $m\approx m_B+m_\Lambda \approx 6.4$ GeV. At the same time the
bound state of the $D\Lambda -D\Sigma $ system was not found for any
reasonable values of the potential strength parameters.

In spite of this encouraging result the $B\Lambda -B\Sigma $ bound state
seems not to be observable as a resonance structure. Indeed, the hadronic
composition of this object when translated into the quark content reads $(b%
\bar q)-(sqq)$. It implies that the found baryonic state can recombine into
the tightly bound heavy baryon $(bqq)$ (or its strange counterpart $(bsq)$)
with the emission of mesons built of light quarks. The mass of the lightest
bottom-flavored baryon $\Lambda _b$ is about 5.75 GeV \cite{pdg}. Because of
large energy release (about 700 MeV) the width of such a decay will be
large, and any resonance-like structure will be completely smeared off. A
state with the genuine pentaquark quantum numbers ( i.e. the $(\bar bsqqq)$
configuration) would correspond to a (loosely bound) state in the $\bar B%
\Lambda -\bar B\Sigma $ system. However, due to negative $G$-parity of the $%
\pi $-meson the one-pion-exchange potential in such a system will have the
sign opposite to that in the $B\Lambda -B\Sigma $ system. Thus the
considered mechanism will generate repulsion and apparently no bound $\bar B%
\Lambda -\bar B\Sigma $ state will exist.

\section{Conclusion}

We have considered possible existence of a heavy-flavored pentaquark with
strangeness $P_{\bar Qs}\equiv (\bar Qsqqq)$. As it was stated in the
Introduction such a multiquark state definitely does not exist as a system
of 5 quarks bound by the color QCD forces. In the approach where a
pentaquark is considered as a bound state of a soliton (nucleon) and a heavy
meson the problem has not been explored. Two circumstances should be taken
into account. First, even in the nonstrange sector this model does not yield
any decisive conclusions. Second, the soliton-based model is ill-suited to
the investigation of the $P_{\bar Qs}$ pentaquark. Indeed, in this model the
hyperon treated as a strange counterpart of the nucleon is already
considered as a bound state of the Skyrme-type soliton (nucleon) and the $K$%
-meson \cite{callan}. Then the consistent description of the heavy meson and
the hyperon interaction would imply the investigation of a three-body system
comprising the heavy meson, the $K$-meson and the nucleon (soliton).
Alternative soliton-based approach where the pentaquark is a bound state of
the nucleon and the (isoscalar) $B_s$ meson would imply that the long-range
one-pion exchange drops out and the interaction is controlled by the forces
of smaller range with the entailing ambiguities related to the
short-distance structure of the soliton.

We have investigated the existence of a molecular-type pentaquark. One-pion
exchange proves to be strong enough to produce a loosely bound state in the $%
B\Lambda -B\Sigma $ system with the quantum numbers $J^\pi (T)=1/2^{-}(1/2)$%
. However, because of the quark content $(b\bar q)-(sqq)$ of this bound
state it can decay rapidly into a bottom-flavored baryon $\Lambda _b$ or its
strange analogue with the emission of light-quark mesons. Because of the
large energy release the width of such a decay will be rather large
precluding possible observation of a resonance-like structure. Thus we can
conclude that the investigation of the heavy-flavored strange pentaquark $P_{%
\bar Qs}$ in various theoretical approaches favor the conclusion that such
an exotic multiquark state is either non existing or is unobservable.
\acknowledgements
The author is indebted to N.T\"ornqvist for discussions.

\newpage
\figure{Dependence of the binding energy $E_b$ in the $\Lambda
H-\Sigma H$ system with the $J^\pi (T)=1/2^{-}(0)$ quantum numbers upon the
formfactor cut-off $\Lambda $ and the potential parameter $g$. Curve 1(2)
corresponds to $\Lambda =1500$ (1200)\ MeV.}
\newpage
\setlength{\unitlength}{0.240900pt}
\ifx\plotpoint\undefined\newsavebox{\plotpoint}\fi
\begin{picture}(1500,900)(0,0)
\font\gnuplot=cmr10 at 10pt
\gnuplot
\sbox{\plotpoint}{\rule[-0.200pt]{0.400pt}{0.400pt}}%
\put(220.0,113.0){\rule[-0.200pt]{292.934pt}{0.400pt}}
\put(220.0,113.0){\rule[-0.200pt]{4.818pt}{0.400pt}}
\put(198,113){\makebox(0,0)[r]{0}}
\put(1416.0,113.0){\rule[-0.200pt]{4.818pt}{0.400pt}}
\put(220.0,257.0){\rule[-0.200pt]{4.818pt}{0.400pt}}
\put(198,257){\makebox(0,0)[r]{2}}
\put(1416.0,257.0){\rule[-0.200pt]{4.818pt}{0.400pt}}
\put(220.0,401.0){\rule[-0.200pt]{4.818pt}{0.400pt}}
\put(198,401){\makebox(0,0)[r]{4}}
\put(1416.0,401.0){\rule[-0.200pt]{4.818pt}{0.400pt}}
\put(220.0,544.0){\rule[-0.200pt]{4.818pt}{0.400pt}}
\put(198,544){\makebox(0,0)[r]{6}}
\put(1416.0,544.0){\rule[-0.200pt]{4.818pt}{0.400pt}}
\put(220.0,688.0){\rule[-0.200pt]{4.818pt}{0.400pt}}
\put(198,688){\makebox(0,0)[r]{8}}
\put(1416.0,688.0){\rule[-0.200pt]{4.818pt}{0.400pt}}
\put(220.0,832.0){\rule[-0.200pt]{4.818pt}{0.400pt}}
\put(198,832){\makebox(0,0)[r]{10}}
\put(1416.0,832.0){\rule[-0.200pt]{4.818pt}{0.400pt}}
\put(220.0,113.0){\rule[-0.200pt]{0.400pt}{4.818pt}}
\put(220,68){\makebox(0,0){0.7}}
\put(220.0,812.0){\rule[-0.200pt]{0.400pt}{4.818pt}}
\put(524.0,113.0){\rule[-0.200pt]{0.400pt}{4.818pt}}
\put(524,68){\makebox(0,0){0.75}}
\put(524.0,812.0){\rule[-0.200pt]{0.400pt}{4.818pt}}
\put(828.0,113.0){\rule[-0.200pt]{0.400pt}{4.818pt}}
\put(828,68){\makebox(0,0){0.8}}
\put(828.0,812.0){\rule[-0.200pt]{0.400pt}{4.818pt}}
\put(1132.0,113.0){\rule[-0.200pt]{0.400pt}{4.818pt}}
\put(1132,68){\makebox(0,0){0.85}}
\put(1132.0,812.0){\rule[-0.200pt]{0.400pt}{4.818pt}}
\put(1436.0,113.0){\rule[-0.200pt]{0.400pt}{4.818pt}}
\put(1436,68){\makebox(0,0){0.9}}
\put(1436.0,812.0){\rule[-0.200pt]{0.400pt}{4.818pt}}
\put(220.0,113.0){\rule[-0.200pt]{292.934pt}{0.400pt}}
\put(1436.0,113.0){\rule[-0.200pt]{0.400pt}{173.207pt}}
\put(220.0,832.0){\rule[-0.200pt]{292.934pt}{0.400pt}}
\put(45,472){\makebox(0,0){$E_b$ [MeV]}}
\put(828,23){\makebox(0,0){g}}
\put(402,688){\makebox(0,0)[l]{1}}
\put(1284,688){\makebox(0,0)[l]{2}}
\put(402,688){\makebox(0,0)[l]{1}}
\put(1284,688){\makebox(0,0)[l]{2}}
\put(220.0,113.0){\rule[-0.200pt]{0.400pt}{173.207pt}}
\put(1058,131){\usebox{\plotpoint}}
\multiput(1058.58,131.00)(0.499,0.902){127}{\rule{0.120pt}{0.820pt}}
\multiput(1057.17,131.00)(65.000,115.298){2}{\rule{0.400pt}{0.410pt}}
\multiput(1123.58,248.00)(0.497,1.552){49}{\rule{0.120pt}{1.331pt}}
\multiput(1122.17,248.00)(26.000,77.238){2}{\rule{0.400pt}{0.665pt}}
\multiput(1149.58,328.00)(0.498,1.840){73}{\rule{0.120pt}{1.563pt}}
\multiput(1148.17,328.00)(38.000,135.756){2}{\rule{0.400pt}{0.782pt}}
\multiput(1187.58,467.00)(0.497,2.158){61}{\rule{0.120pt}{1.812pt}}
\multiput(1186.17,467.00)(32.000,133.238){2}{\rule{0.400pt}{0.906pt}}
\multiput(1219.58,604.00)(0.497,2.442){61}{\rule{0.120pt}{2.038pt}}
\multiput(1218.17,604.00)(32.000,150.771){2}{\rule{0.400pt}{1.019pt}}
\multiput(1251.58,759.00)(0.492,2.693){21}{\rule{0.119pt}{2.200pt}}
\multiput(1250.17,759.00)(12.000,58.434){2}{\rule{0.400pt}{1.100pt}}
\put(1263.17,822){\rule{0.400pt}{2.100pt}}
\multiput(1262.17,822.00)(2.000,5.641){2}{\rule{0.400pt}{1.050pt}}
\put(265,142){\usebox{\plotpoint}}
\multiput(265.58,142.00)(0.494,1.386){27}{\rule{0.119pt}{1.193pt}}
\multiput(264.17,142.00)(15.000,38.523){2}{\rule{0.400pt}{0.597pt}}
\multiput(280.58,183.00)(0.494,1.650){29}{\rule{0.119pt}{1.400pt}}
\multiput(279.17,183.00)(16.000,49.094){2}{\rule{0.400pt}{0.700pt}}
\multiput(296.58,235.00)(0.494,2.105){27}{\rule{0.119pt}{1.753pt}}
\multiput(295.17,235.00)(15.000,58.361){2}{\rule{0.400pt}{0.877pt}}
\multiput(311.58,297.00)(0.494,2.447){27}{\rule{0.119pt}{2.020pt}}
\multiput(310.17,297.00)(15.000,67.807){2}{\rule{0.400pt}{1.010pt}}
\multiput(326.58,369.00)(0.497,2.783){59}{\rule{0.120pt}{2.306pt}}
\multiput(325.17,369.00)(31.000,166.213){2}{\rule{0.400pt}{1.153pt}}
\multiput(357.58,540.00)(0.498,3.446){71}{\rule{0.120pt}{2.835pt}}
\multiput(356.17,540.00)(37.000,247.116){2}{\rule{0.400pt}{1.418pt}}
\multiput(394.59,793.00)(0.482,3.474){9}{\rule{0.116pt}{2.700pt}}
\multiput(393.17,793.00)(6.000,33.396){2}{\rule{0.400pt}{1.350pt}}
\end{picture}

\begin{references}
\bibitem{richard}  C.Gignoux {\it et al}., Phys.Lett. {\bf B193} (1987) 323
\bibitem{lipkin}  H.J.Lipkin, Phys.Lett. {\bf B195} (1987) 484
\bibitem{fleck}  S.Fleck {\it et al}., Phys.Lett. {\bf B220} (1989) 616
\bibitem{zouzou}  S.Zouzou and J.-M.Richard, Few-Body Systems, {\bf 16 }%
(1994) 1
\bibitem{riska}  D.O.Riska and N.N.Scoccola, Phys.Lett. {\bf B299} (1993) 338
\bibitem{sh}  M.Shmatikov, Phys.Lett. {\bf B349} (1995) 411
\bibitem{Oh}  Y.Oh, B.-Y.Park and D.-P.Min, Phys.Lett. {\bf B331} (1994) 362
\bibitem{Nij}  M.M.Nagels, T.A.Rijken and J.J.de Swart, Phys.Rev. {\bf D12}
(1975) 744
\bibitem{To}  N.A.T\"ornqvist, Phys.Rev.Lett. {\bf 67} (1992) 556
\bibitem{Blatt}  J.M.Blatt and V.F.Weisskopf, Theoretical nuclear physics (%
{\it Wiley}, New York, 1952)
\bibitem{pdg}  Particle Data Group, Phys.Rev. {\bf D50} (1994) 1173
\bibitem{callan}  C.D.Callan, K.Hornbostel and I.Klebanov, Phys.Lett. {\bf %
B202} (1988) 269
\end{references}
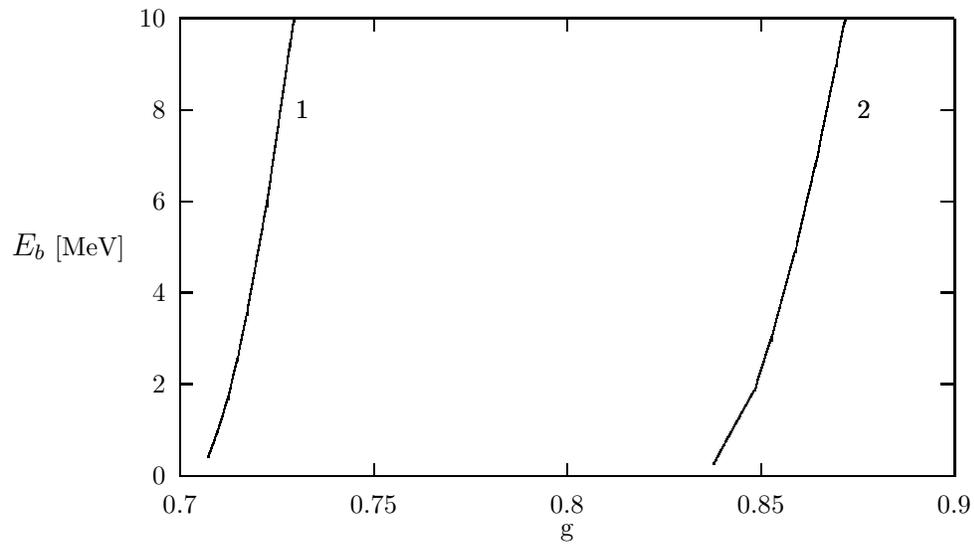
\end{document}